\documentclass[twocolumn,showpacs,preprintnumbers,amsmath,amssymb,aps]{revtex4}
\usepackage{graphicx}
\begin{document}
\title {Superconductor-semiconductor magnetic microswitch}

\author{C. Castellana}
\author{F. Giazotto}
\email{f.giazotto@sns.it}
\author{M. Governale}
\author{F. Taddei}
\author{F. Beltram}
\affiliation{NEST CNR-INFM and Scuola Normale Superiore, I-56126 Pisa, Italy}

\begin{abstract}
%change
A hybrid superconductor--two-dimensional electron gas microdevice is presented. Its working principle is based on the suppression of Andreev reflection at the superconductor-semiconductor interface caused by a magnetic barrier generated by a ferromagnetic strip placed on top of the structure. Device switching is predicted with fields up to some mT and working frequencies of several GHz, making it promising for applications ranging from microswitches and storage cells to magnetic field discriminators.
\end{abstract}


\maketitle

Normal metal-superconductor (NS) junctions show peculiar low-voltage transport properties due to the presence of the superconducting gap that can be exploited for several electronic applications \cite{ruggiero} ranging from
superconducting interferometry \cite{eom2} and nonvolatile storage-cell engineering \cite{johnson1,johnson2,johnson3}
to microrefrigeration \cite{cool0} and Josephson-effect devices \cite{likharev}.
In these systems electronic transport is mediated by Andreev reflection \cite{andreev} for subgap voltages. This is a scattering process occurring at the NS interface consisting of the coherent evolution of an electron into a retro-reflected hole thus describing the injection of a Cooper pair into the condensate.
In the tunneling regime and at low temperatures subgap conductance is drastically suppressed with respect to the normal-state. Subgap conductance
can therefore be effectively controlled by tuning Andreev reflection. This is of interest not only from the fundamental physics point of view, but also for its potential impact on nanoelectronics applications.     

In this Letter we propose a hybrid superconductor-semiconductor device whose 
subgap conductance can be controlled by changing the strength of a magnetic barrier induced by a ferromagnetic strip. The simplicity of the structure together with the mature superconductor-semiconductor junction technology \cite{batov} make this system promising for a number of electronic applications.

\begin{figure}[t!]
\begin{center}
\includegraphics[width=\columnwidth]{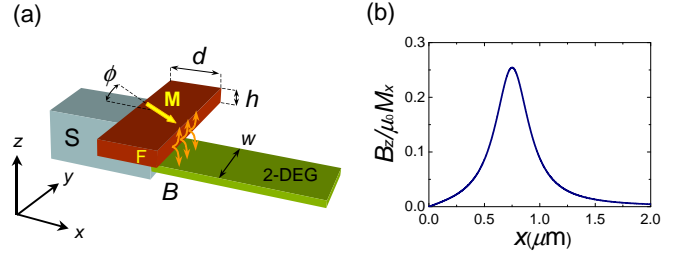}
\end{center}
\caption{(Color online) (a) Schematic view of the proposed device. 
The superconductor-2DEG junction conductance is controlled by the fringe field generated by a ferromagnetic strip placed on top of the structure. The strength of the magnetic barrier in the two-dimensional electron gas can be varied by changing the orientation of the magnetization ($\mathbf{M}$).
(b) Spatial profile of the $B_z$-component of the magnetic fringe field for $z_0=185$~nm, $h=300$~nm, and $d=1.5~\mu$m. The transport direction is $x$.  }
\label{fig:fringe}
\end{figure}
A sketch of our device is shown in Fig. \ref{fig:fringe}(a).
It consists of a superconductor--two-dimensional electron gas (2DEG) ballistic junction. 
The superconducting interface is located at $x=0$, and transport occurs along the $x$ axis.  On top of the junction a ferromagnetic strip of width $d$ and thickness $h$ is deposited. It has a homogeneous magnetization $\mathbf{M}$ 
and occupies the region defined by
$-d/2\le x \le d/2$ and $-h/2\le z \le h/2$ (note that the origin of the $z$ axis is 
at the mid-height of the ferromagnetic strip and $z_0$ is the position of the 2DEG plane). 
Electron transport in the 2DEG is affected by the perpendicular magnetic field 
generated  by the $x$-component of the magnetization $M_x=|\mathbf{M}| \cos\phi$. 
The $z$-component of the magnetic field has the profile of a double magnetic barrier \cite{matulis,govboese} whose strength can be tuned by 
rotating the magnetization of the ferromagnet (hence, by varying $M_x$). 
The $z$-component of the magnetic field in the 2DEG plane, in the limit $h\ll z_0$ and $h\ll d$, is given by \cite{matulis}:
$B_{z}(x)=\frac{\mu_0}{2 \pi} M_{x}\frac{h}{d}\Big(K(x+d/2,z_0)-K(x-d/2,z_0)\Big)\theta(x)$, 
with $K(x,z)=-zd/(x^{2}+z^{2})$, $\theta(x)$ the step function, and 
$\mu_0$ the magnetic permeability of vacuum. 
An example of the magnetic barrier profile is shown in Fig.~\ref{fig:fringe}(b). 
Note that one of the peaks is located deep in the superconductor region and does not affect the transport properties of the interface.
Although the above expression for the 
magnetic field has been derived for  $h\ll z_0$, it is still a very good 
approximation even when $h$ and $z_0$ are of the same order \cite{govboese}, and we shall use it throughout this Letter. 
If the energy of quasiparticles contributing to transport is smaller than the superconducting gap, transport is mediated 
by Andreev reflection. As will be shown in the following, the presence of the 
magnetic barrier can \emph{suppress} Andreev reflection, and hence the subgap conductance. 
A setup similar to the one presented here was proposed and realized in the context of all-metal superconducting weak links \cite{eom2,johnson1,johnson2,johnson3}. The working principle, however, was different and in those structures the magnetic fringe field was used to locally quench superconductivity.

In order to study electron transport in the structure we make use of the the Bogolubov-de Gennes equation \cite{bdg}:
\begin{equation}
 \begin{pmatrix}
  H_{0}^{\sigma} &  \Delta(x)\\
  \Delta(x)        & -{H_{0}^{\sigma}}^{*}
 \end{pmatrix}
 \begin{pmatrix}
  u^{\sigma}\\
  v^{-\sigma} 
 \end{pmatrix}
 = E
 \begin{pmatrix}
  u^{\sigma}\\
  v^{-\sigma}
 \end{pmatrix},
\label{eq:ehSm}
\end{equation}
with 
\begin{equation}
H_{0}^{\sigma}=(\mathbf{p}-e\mathbf{A}(x)) \frac{1}{2m(x)}(\mathbf{p}-e\mathbf{A}(x))+V_{\sigma}(x)+U(x)-\epsilon_{\text{F,S}},
\end{equation}
where $\mathbf{p}=-i\hbar \vec{\nabla}$, $\mathbf{A}(x)=A_{y}(x)\hat{y}$ is the vector potential in the London gauge, $e$ the electron charge, $u$ and $v$ are the coherence factors, and $\sigma=\pm 1$ is the spin. The excitation energy $E$ is measured from the condensate chemical potential $\epsilon_{\text{F,S}}$.
The potential $U$ describes the subband-bottom mismatch between the superconductor and the 2DEG.
Zeeman splitting $V_{\sigma}(x)$ is given by
$ V_{\sigma}(x)=\frac{1}{2}\sigma g_{\text{2DEG}}\mu_{B}B(x)$, where $\mu_{B}$
is the Bohr magneton, and $g_{\text{2DEG}}$ the effective g-factor. 
The electron mass and the pairing potential are given, respectively, by 
$m(x)=m_{\text{2DEG}}\theta(x)+m_{0}\theta(-x)$ and
$\Delta(x)=\Delta \theta(-x)$. 

Within the Landauer-Buttiker scattering approach \cite{Landauer} the  
junction finite-temperature differential conductance is given by:
\begin{eqnarray}
G(V) & = & \frac{e^2}{h}\int_{-\infty}^{+\infty}dE \sum_{\sigma}\Big[ N^{+\sigma}(E)-R_{0}^{+\sigma}(E)+ \nonumber \\         &  & R_{a}^{+\sigma}(E)\Big] \Big(  -\frac{\partial f(E-eV)}{\partial E} \Big), 
\end{eqnarray}
where $R_{0}^{+\sigma}$ ($R_{a}^{+\sigma}$) is the normal (Andreev) reflection probability for spin-$\sigma$ quasiparticles, $N^{+,\sigma}$ is the number of open electron channels in the junction, and $f$ is the equilibrium Fermi distribution at temperature $T$.
The scattering amplitudes are numerically evaluated by a recursive Green's function technique within a tight-binding description of the system \cite{sanvito}. 
In order to simulate a realistic S-2DEG structure we assumed 
$w=0.5$ $\mu$m, $d=1.5$ $\mu$m, $z_0=185$ nm, $h=300$ nm, and $g_{\text{2DEG}}=-20$. 
Within the tight-binding scheme, $H_0^{\sigma}$ is a matrix characterized by on-site energies at each site $i$ ($\epsilon_{0}^{i}$) and hopping potentials between nearest-neighbors sites ($\gamma_{ij}$).
In particular, in order to describe the different band structures in S and  the 2DEG, we have taken $\epsilon_{0}^{\text{2DEG}}=15.97$ eV, $\epsilon_{0}^{\text{S}}=2.4$ eV, $\gamma^{\text{2DEG}}=4.0$ eV, and $\gamma^{\text{S}}=1.0$ eV. The parameters chosen for the semiconductor region are suitable for an InAs \cite{giazotto} or In$_{0.75}$Ga$_{0.25}$As \cite{capotondi} 2DEG  with charge density $\simeq 4.4\times10^{11}$ cm$^{-2}$, and  effective mass $m_{\text{2DEG}}=0.035~m_{0}$. 
$\Delta$ is a diagonal matrix with non-zero and constant elements ($\Delta_0$) only in S.
$\Delta_{0}$ is assumed to follow the BCS temperature dependence, and we chose aluminum (Al) as the superconducting electrode ($\Delta_0(T=0)=180~\mu$eV).
With the parameters given above the junction normal-state resistance for $B_z =0$ is $R_{N}\sim 3.2$ k$\Omega$.

\begin{figure}[t!]
\begin{center}
\includegraphics[width=\columnwidth]{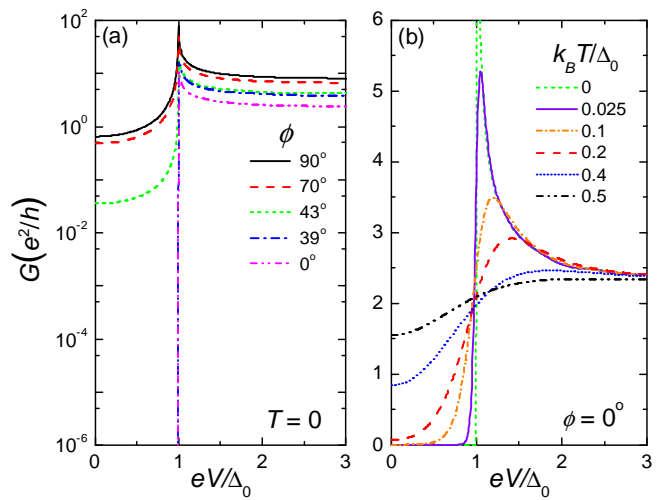}
\end{center}
\caption{(Color online)
(a) Differential conductance $G$ vs bias voltage  at $T=0$ and $\mu_{0}M=1.8$ T for several angles $\phi$. 
(b) $G$ vs bias voltage for several temperatures $T$ at $\mu_{0}M=1.8$ T and $\phi =0^{\circ}$.} 
\label{fig:BmaxT}
\end{figure}

The results of our calculations are shown in 
Fig. \ref{fig:BmaxT} for $\mu_0 M=1.8$ T (i.e., the saturation magnetization typical of Co). 
Figure \ref{fig:BmaxT}(a) shows $G(V)$ at $T=0$ for different angles $\phi$.
As expected, the conductance shows very different behavior for $|eV|<\Delta_0$ and $|eV|>\Delta_0$. 
When the peak of the fringe field grows by lowering the magnetization angle, the subgap conductance is drastically suppressed (down to zero within the numerical error). The normal-state conductance is much less affected and is reduced at most by a factor $\sim 3$ at $\phi =0^{\circ}$. For magnetic barrier strengths exceeding the critical value (with the chosen parameters, around $\phi =39 ^{\circ}$) the S-2DEG contact thus behaves like an ideal superconducting \textit{tunnel} junction showing a subgap conductance several orders of magnitude smaller than in the normal state \cite{tinkham}.
By increasing the temperature (see Fig. \ref{fig:BmaxT}(b)) the junction $G(V)$ characteristic resembles that of an S-insulator-N tunnel contact, i.e., 
the subgap conductance increases, reaching the normal-state value at the superconductor critical temperature.

\begin{figure}[t!]
\begin{center}
\includegraphics[width=\columnwidth]{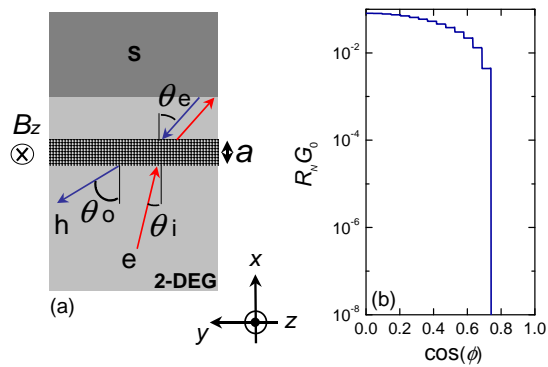}
\end{center}
\caption{(Color online) (a) Semiclassical picture of the Andreev reflection suppression mechanism caused by a magnetic barrier. $e$ and $h$ represent electron- and hole-like quasiparticles, respectively.
(b) Normalized zero-bias differential conductance $R_N G_{0}$ vs $\text{cos}(\phi)$ at $T=0$ and $\mu_0 M=1.8$~T.} 
\label{fig:HZR}
\end{figure}

The physical origin of this suppression of Andreev reflection can be easily understood within a semiclassical picture (see Fig. \ref{fig:HZR}(a)). For simplicity let us consider 
vanishing bias voltage $V$ and temperature, so that transport occurs at the Fermi energy. 
Furthermore, we assume a magnetic field such that  
$\mathbf{B}(x)= B\hat{z}$ for $0\leq x\leq a$, and $\mathbf{B}(x)=0$ elsewhere. 
Let us consider an electron (e) propagating in the $xy$ plane 
and impinging on the magnetic barrier at an angle $\theta_{i}$. 
When the cyclotron radius is larger than $a$, the electron leaves the barrier at an angle $\theta_{e}$, and it is reflected as a hole (h) at the S-N interface (within this simplified description we can neglect normal reflection). 
After crossing the magnetic barrier, the hole propagates at an angle $\theta_{o}$ related to $\theta_{i}$ by
$\sin \theta_{o}-\sin \theta_{i}=\frac{2a\omega_\text{c}}{v_{\text{F}}}$,
where $v_{\text{F}}$ is the Fermi velocity in the 2DEG, and $\omega_{\text{c}}=|e B|/
m_{\text{2DEG}}$. In full analogy with total internal reflection the hole is prevented from crossing the magnetic barrier if $\theta_{i}$ exceeds the critical value $(-1)^{\text{sign}(eB)}\sin \theta_{i}^{c}=-1+\frac{2a \omega_{\text{c}}}{v_{\text{F}}}$. On the other hand
Andreev reflection is suppressed if the incident angle reaches 
$-\text{sign}(e B) \pi/2$.
Hence, the critical field value ($B^{\text{c}}$) leading to the suppression of Andreev reflection satisfies the relation
\begin{equation}
\frac{a|e B^{c}|}{m_\text{2DEG} v_{\text{F}}} = 1.
\label{eq:critcond}
\end{equation}
Here it is important to point out that, for a s-wave superconductor, interface roughness, which occurs on the atomic length scale, will virtually have no effect on Andreev reflection \cite{andreev,roughness}, the latter being determined by electron-hole correlations which extend on a much larger length scale.

By controlling the intensity of the magnetic barrier it is thus possible to have access to different transport regimes in the \textit{same} structure. We can tune transport from that typical of a relatively transparent junction to an exponentially suppressed subgap conductance, a behavior typical of a tunnel junction. The main difference between the present approach and a conventional barrier, e.g., a thin oxide layer or a barrier produced by electrostatic gating, is that the fringe field strongly affects the contact subgap conductance in the superconducting state, but marginally alters the junction normal-state resistance \cite{wire}. 
The full switching behavior of the junction is shown in Fig. \ref{fig:HZR}(b) which displays the normalized zero-bias differential conductance ($R_N G_0$) vs $\text{cos}(\phi)$ at $T=0$ and $\mu_0 M=1.8$~T.
By reducing the magnetization angle,
$G_{0}$ is suppressed by many orders of magnitudes with respect to $R_N^{-1}$ (the steps appearing in $G(\phi)$ are due to closing of transport channels induced by the magnetic barrier).

A first obvious application of the proposed structure is the implementation of a magnetic \emph{switch},  by taking advantage of the narrow transition in the $G_{0}(\phi)$ curve and of the large $(R_N G_0)^{-1}$ value (see Fig. 3(b)). Additionally, the system could be operated as a magnetic field \emph{discriminator}. In such a configuration, the junction should be set near the transition point with a properly chosen fringe field (i.e., at a suitable angle $\phi$); then, any additional external magnetic field would trigger the device to switch.
Also the implementation of \emph{nonvolatile} storage cells can be easily envisioned: $\mathbf{M}$ can be rotated from $\phi=0$ to $\phi=\pi/2$ through metallic write wires, along the lines of Refs. \cite{eom2,johnson1,johnson2,johnson3}, while the junction state is maintained with zero applied power by the remnant magnetization and is nonvolatile, even at temperatures larger than the critical temperature of the superconductor. 

We should like to summarize the essential requirements toward a realistic 
implementation of the device:
(i) a \textit{semiconductor}-based hybrid structure is required. It allows to overcome the difficulty of achieving complete Andreev suppression with an \textit{all-metal} system (from Eq. (\ref{eq:critcond}) it follows that unrealistically-high fringe fields would be required to suppress the Andreev reflection with normal metals);
(ii) any Schottky barrier at the S-2DEG contact should be avoided, in order to minimize the junction normal-state resistance.
In light of these considerations, 2DEGs made in the  InAs \cite{giazotto} or In$_{x}$Ga$_{1-x}$As (with $x\geq 0.75$) \cite{capotondi} systems are ideal candidates for the Sm region, and Al or Nb could be used for the S electrode.
As far as the ferromagnetic element is concerned, a  Permalloy (Ni$_{0.8}$Fe$_{0.2}$) \cite{johnson1,johnson2}, Co or Co$_{90}$Fe$_{10}$ \cite{eom2,johnson3} thin layer could provide magnetizations ($\mu_{0}M$) ranging from $1$ to $2$ T. Furthermore, $\mathbf{M}$ can be rotated by externally applying an in-plane static magnetic field as low as some $10^{-3}$ T \cite{eom2,johnson1,johnson2,johnson3}, while
its rotation frequency (determining the device speed) might be pushed up to several GHz allowing, in principle, fast operation \cite{gerrits}.

In conclusion, we proposed a superconductor-2DEG hybrid microstructure that exploits the magnetic fringe field generated by a ferromagnetic strip to control the Andreev reflection at the superconductor--semiconductor interface. This system shows a high potential for device applications ranging from fast switches and magnetic field discriminators to nonvolatile storage cells.

Partial financial support from MIUR under FIRB program RBNEO1FSWY is gratefully acknowledged.

%------------------------------------------ References


\begin{thebibliography}{99}

\bibitem{ruggiero}
				\emph{Superconducting Devices}, edited by S. Ruggiero and D. Rudman (Academic, Boston, 1990).

\bibitem{eom2}
				J. Eom and M. Johnson, Appl. Phys. Lett. \textbf{79}, 2486 (2001).

\bibitem{johnson1}
  			T. W. Clinton and M. Johnson, Appl. Phys. Lett. \textbf{70}, 1170 (1997).

\bibitem{johnson2}
  			T. W. Clinton and M. Johnson, Appl. Phys. Lett. \textbf{76}, 2116 (2000).

\bibitem{johnson3}
  			T. W. Clinton, P. R. Broussard, and M. Johnson, J. Appl. Phys. \textbf{91}, 1371 (2001).

\bibitem{cool0}
  			J. P. Pekola, T. T. Heikkil\"{a}, A. M. Savin, J. T. Flyktman, F. Giazotto, and F. W. J. Hekking, Phys. Rev. Lett. 								\textbf{92}, 056804 (2004).

\bibitem{likharev}
				K. K. Likharev, Rev. Mod. Phys. \textbf{51}, 101 (1979).

\bibitem{andreev} A.F. Andreev, Zh. Eksp. Teor. Fiz. {\bf 46}, 1823 (1964); [Sov. Phys. JETP {\bf 19}, 1228 (1964)].

\bibitem{batov}
				I. E. Batov, Th. Sch\"{a}pers, A. A. Golubov, and A. V. Ustinov, J. Appl. Phys. \textbf{96}, 3366 (2004).

\bibitem{matulis}
  			A. Matulis, F. M. Peeters, and P. Vasilopoulos, Phys. Rev. Lett. \textbf{72}, 1518 (1994).

\bibitem{govboese} M. Governale and D.  Boese, Appl. Phys. Lett. \textbf{77}, 3215 (2000).

\bibitem{bdg}
				P. G. de Gennes, \emph{Superconductivity of Metals and Alloys} (Addison-Wesley, New York, 1989).

\bibitem{Landauer}
  				R. Landauer, Philos. Mag. \textbf{21}, 863 (1970); M. Buttiker, Phys. Rev. Lett. \textbf{57}, 1761 (1986).
  
\bibitem{sanvito}
  			S. Sanvito, C. J. Lambert, J. H. Jefferson, and A. M. Bratkovsky, Phys. Rev. B \textbf{59}, 11936 (1999).

\bibitem{giazotto}
				F. Giazotto, K. Grove-Rasmussen, R. Fazio, F. Beltram, E. H. Linfield, and D. A. Ritchie, J. Supercond. \textbf{17}, 317 (2004).
				
\bibitem{capotondi}
				F. Capotondi, G. Biasiol, I. Vobornik, L. Sorba, F. Giazotto, A. Cavallini, and F. Fraboni, J. Vac. Sci. Technol. B 											\textbf{22}, 702 (2004).

\bibitem{tinkham}
				M. Tinkham, \emph{Introduction to Superconductivity} (McGraw-Hill, New York, 1996).

\bibitem{roughness}	Y. Nagato, and K. Nagai, Phys. Rev. B \textbf{69}, 104507 (2004); P. A. M. Benistant, H. van Kempen, and P. Wyder, Phys. Rev. Lett. \textbf{51}, 817 (1983); P. A. M. Benistant, A. P. van Gelder, H. van Kempen, and P. Wyder, Phys. Rev. B \textbf{32}, 3351 (1985).

\bibitem{wire}
				We note that similar fringe fields could also be generated  by a (super)conducting strip with a current driven through it \cite{matulis}. However, the current necessary to produce the required magnetic barrier strength would largely exceed the superconductor critical current, thus leading to large power dissipation.
				
\bibitem{gerrits}
  			Th. Gerrits, H. A. M. van den Berg, J. Hohlfeld, L. Bar, and Th. Raising, Nature \textbf{418}, 509 (2002).

\end{thebibliography}
\end{document}